\documentclass[journal]{IEEEtran}
\usepackage{xcolor}
\usepackage{arydshln}
\usepackage{amsmath}
\usepackage{amssymb}
\usepackage{orcidlink}

\usepackage{booktabs}
\usepackage{multirow}
\usepackage{makecell}
\usepackage{nicematrix}
\usepackage{graphicx}
\usepackage{subfig}
\usepackage{float}
\usepackage{algorithm,algorithmic}
\usepackage{xcolor}
\usepackage{colortbl}
\usepackage[final]{review}
\definecolor{graybackground}{gray}{0.9}

\DeclareMathOperator*{\argmax}{arg\,max}
\begin{document}

\title{VINP: Variational Bayesian Inference with Neural Speech Prior for Joint ASR-Effective Speech Dereverberation and Blind RIR Identification}

\author{Pengyu~Wang\orcidlink{0000-0001-5768-0658}\thanks{Pengyu Wang is with Zhejiang University and also with Westlake University, Hangzhou, China (e-mail: wangpengyu@westlake.edu.cn).},
Ying Fang\orcidlink{0009-0003-8767-1172}\thanks{Ying Fang is with Zhejiang University and also with Westlake University, Hangzhou, China (e-mail: fangying@westlake.edu.cn).},
and
Xiaofei~Li\orcidlink{0000-0003-0393-9905}
\thanks{Xiaofei Li is with the School of Engineering, Westlake University, and the Institute of Advanced Technology, Westlake Institute for Advanced Study, Hangzhou, China (e-mail: lixiaofei@westlake.edu.cn). }
\thanks{Xiaofei Li: corresponding author.}}


\maketitle

\begin{abstract}
Reverberant speech, denoting the speech signal degraded by reverberation, contains crucial knowledge of both anechoic source speech and room impulse response (RIR).
This work proposes a variational Bayesian inference (VBI) framework with neural speech prior (VINP) for joint speech dereverberation and blind RIR identification.
In VINP, a probabilistic signal model is constructed in the time-frequency (T-F) domain based on convolution transfer function (CTF) approximation. 
For the first time, we propose using an arbitrary discriminative dereverberation deep neural network (DNN) to estimate the prior distribution of anechoic speech within a probabilistic model. 
By integrating both reverberant speech and the anechoic speech prior, VINP yields the maximum a posteriori (MAP) and maximum likelihood (ML) estimations of the anechoic speech spectrum and CTF filter, respectively.
After simple transformations, the waveforms of anechoic speech and RIR are estimated.
VINP is effective for automatic speech recognition (ASR) systems, which sets it apart from most deep learning (DL)-based single-channel dereverberation approaches.
Experiments on single-channel speech dereverberation demonstrate that VINP attains state-of-the-art (SOTA) performance in mean opinion score (MOS) and word error rate (WER).
For blind RIR identification, experiments demonstrate that VINP achieves SOTA performance in estimating reverberation time at 60 dB (RT60) and advanced performance in direct-to-reverberation ratio (DRR) estimation.
Codes and audio samples are available online\footnote{\url{https://github.com/Audio-WestlakeU/VINP}}.

\end{abstract}


\IEEEpeerreviewmaketitle

\section{Introduction}

\IEEEPARstart{R}{everberation}, which is formed by the superposition of multiple reflections, scattering, and attenuation of sound waves in a closed space, is one of the main factors degrading speech quality and intelligibility in daily life.
The reverberant speech contains the crucial knowledge of both anechoic source speech and room impulse response (RIR).
The anechoic source speech is clean and serves as the target of the dereverberation task (with possible time delay and scaling).
Moreover, RIR characterizes the sound propagation in an enclosure.
\addnote[rirapplications]{1}{An accurate estimate of RIRs contributes to applications including speech recognition~\cite{ratnarajah2023towards}, speech enhancement~\cite{vincent2018audio} and augmented/virtual reality~\cite{schissler2014high}.}
Therefore, given a reverberant microphone recording, there is a strong need to estimate the anechoic speech and RIR, leading to two distinct tasks: speech dereverberation and blind RIR identification, respectively.

A series of classical speech dereverberation approaches build deterministic signal models of anechoic source speech, RIR, and reverberant microphone recording.
After that, the task is solved by designing an inverse filter in the time domain~\cite{nakatani2010speech} or time-frequency (T-F) domain~\cite{nakatani2010speech,nakatani2008blind,kinoshita2009suppression,yoshioka2012generalization,kodrasi2014frequency}.
In contrast, another series of classical methods regard the anechoic source speech and the reverberant microphone recording as random signals, and use hierarchical probabilistic models to describe the generation process of the observation.
After that, speech dereverberation is performed by estimating every unknown part in the probabilistic model, including hidden variables and model parameters~\cite{schmid2014variational,mohammadiha2015speech}.
The construction of the probabilistic model is highly flexible. 
Similar models can be applied to a variety of fields, such as speech denoising~\cite{schmid2014variational,8462460,8492427} and direct-of-arrival estimation~\cite{yang2012off,wang2022off,wang2022joint}.
These classical approaches are model-based, which makes them free from generalization problems.
However, due to the insufficient utilization of prior knowledge regarding speech and reverberation, their performance remains unsatisfactory.

In recent decades, data-driven deep learning (DL)-based approaches have developed rapidly. 
These approaches rely less on the assumptions of signal models, instead directly learn the characteristics of speech signals using deep neural networks (DNNs).
The core research of DL-based approaches focus in the design of DNN structures, features, and loss functions.
The most straightforward DL-based idea is to construct a discriminative DNN to learn the mapping from degraded speech to target speech.
For instance, authors in \cite{han2015learning,luo2018real,zhao2020monaural,hao2021fullsubnet,zhou2023speech,xiong2022spectro} developed various discriminative DNNs and loss functions to build mappings in various feature domains.
Another idea is to consider speech dereverberation as a generative task and utilize generative DNNs, such as variational autoencoder (VAE)~\cite{Kingma2014auto}, generative adversarial network (GAN)~\cite{goodfellow2020generative}, and diffusion model (DM)~\cite{ho2020denoising}, to generate anechoic speech~\cite{fu2019metricgan,abdulatif2024cmgan,richter2023speech,lemercier2023storm}.
Thanks to the powerful non-linear modeling ability of DNNs, DL-based methods are able to make the best of a large amount of data and typically lead to better perceptual performance than classical approaches.
However, when applied as single-channel front-end systems without joint training, these DL-based methods may introduce waveform artifacts that degrade subsequent automatic speech recognition (ASR) performance~\cite{iwamoto22_interspeech,iwamoto2024does}.

Particularly, some approaches combine DNNs and probabilistic signal models, forming a so-called semi-supervised~\cite{mysore2011non,bando2018statistical} or unsupervised~\cite{bie2022unsupervised,baby2021speech} paradigm.
Unlike fully supervised DL-based approaches, these methods train VAEs with only clean speech corpora.
At the inference stage, the VAE decoder participates in solving the probabilistic model by estimating the prior distribution of clean speech.
Usually, the Markov chain Monte Carlo (MCMC) algorithm is used to sample the latent variables in VAE.
Compared with classical methods, the prior generated by VAE has higher quality and can yield better performance.
Such algorithms are applied to both speech denoising~\cite{mysore2011non,bando2018statistical,bie2022unsupervised} and speech dereverberation~\cite{baby2021speech,wang2024rvae}.
For instance, in \cite{baby2021speech}, the authors developed a Monte Carlo expectation-maximization (EM) dereverberation algorithm based on a convolutional VAE and non-negative matrix factorization (NMF).
In our previous work RVAE-EM~\cite{wang2024rvae}, we employed a more powerful recurrent VAE and found that when the anechoic speech prior is of sufficient quality, MCMC is unnecessary, thus avoiding the repeated inference of VAE decoders.
Moreover, by introducing supervised data into the training of VAE, we also obtained a supervised version of RVAE-EM that performs better than the unsupervised one.
This finding inspired our current work. 
Given the availability of many advanced dereverberation DNNs, it is possible to directly apply them as supervised prior estimators in the probabilistic model solution through straightforward modifications.

Within the domain of audio signal processing, blind RIR identification from reverberant microphone recordings is a crucial and challenging area of research.
Currently, 
methods for blind RIR identification are rather scarce. 
In recent years, with the development of DL techniques, some DL-based approaches have been proposed~\cite{steinmetz2021filtered,richard2022deep,lemercier2024unsupervised}.
For instance, in FiNS~\cite{steinmetz2021filtered}, the authors utilized DNN to learn the direct path, early reflection, and late reverberation of RIR separately.
The authors in BUDDy~\cite{lemercier2024unsupervised} established a parameterized model for the reverberation effect and proposed an unsupervised method for joint speech dereverberation and blind RIR estimation.

In this paper, we propose a \textbf{V}ariational Bayesian \textbf{I}nference framework with \textbf{N}eural speech \textbf{P}rior (VINP) for joint speech dereverberation and blind RIR identification.
Our motivation is as follows: 
The generation of reverberant recording can be modeled in a probabilistic model based on convolution transfer function (CTF) approximation~\cite{avargel2007system,talmon2009relative}.
By treating the DNN output as the prior distribution of anechoic speech and combining it with the reverberant recording to solve the probabilistic signal model, we can analytically estimate the anechoic spectrum and the CTF filter, and subsequently obtain the anechoic speech and RIR waveforms.
By doing this, VINP avoids the direct utilization of DNN output but still leverages the powerful nonlinear modeling capability of DNN, and further ensures that the estimates and observations align with the CTF-based signal model, thereby benefiting ASR.
Different from our previous work RVAE-EM~\cite{wang2024rvae}, 
in VINP, we propose to employ an arbitrary discriminative DNN to learn the prior distribution of anechoic speech by modifying the loss function during training.
Moreover, a major drawback of RVAE-EM is that the computational complexity scales cubically with speech duration.
In VINP, we use variational Bayesian inference (VBI)~\cite{beal2003variational,tzikas2008variational} to analytically solve the probabilistic model.
Thanks to VBI, the computational complexity in VINP increases linearly with the speech duration.
Parallel computation is available across T-F bins for efficient implementation.

This paper has the following contributions:
\begin{itemize}
    \item We propose VINP, a novel framework for joint speech dereverberation and blind RIR identification.
    For the first time, we propose introducing an arbitrary discriminative dereverberation DNN as a backbone into VBI to successfully complete these two tasks.
    \item 
    VINP avoids the direct utilization of DNN output but still utilizes the powerful nonlinear modeling capability of the network, and further ensures that the estimates and observations align with the CTF-based signal model.
    As a result,  VINP attains state-of-the-art (SOTA) performance in mean opinion score (MOS) and word error rate (WER).
    \item 
    VINP can be used for blind RIR identification.
    Experiments demonstrate that VINP attains SOTA and advanced levels in the blind estimation of reverberation time at 60dB (RT60) and direct-to-reverberation ratio (DRR). 
    \item From the perspective of computational complexity, VINP exhibits linear scaling with respect to the speech duration, representing a significant improvement over the cubical scaling in our previous work RVAE-EM~\cite{wang2024rvae}.
    Moreover, the VBI procedure enables parallel processing across T-F bins, which further accelerates inference.
\end{itemize}

The remainder of this paper is organized as follows: 
Section \ref{sec:problem_formulation} formulates the signal model and the tasks.
Section \ref{sec:prop} details the proposed VINP framework. 
Experiments and discussions are presented in Section \ref{sec:expset}. 
Finally, Section \ref{sec:conclusion} concludes the entire paper.

\section{Signal Model and Tasks}
\label{sec:problem_formulation}

\subsection{Signal Model}

Considering the scenario of a single static speaker and stationary noise, the reverberant recording (also known as observation) received by a distant microphone can be modeled in the time domain as
\begin{equation}
    \label{eq:signal_model_time}
    x(n)=h(n)*s(n)+w(n),
\end{equation}
where $*$ is the convolution operator, $n$ is the index of sampling points, $x(n)$ is the observation signal, $s(n)$ is the anechoic source speech signal, $h(n)$ is the RIR which describes a time-invariant linear filter, and $w(n)$ is the background additive noise.

\addnote[RIR_long]{1}{Analyzing and processing speech signals and the very long RIR filter (normally thousands of taps) in the time domain poses significant challenges. 
For example, the estimation of filter, e.g. Eq.~(\ref{eq:est_H}) in the proposed method, shares similar solutions in T-F domain and time domain.
But the solution in time domain requires the inversion of a very large correlation matrix, which suffers from very high computational/memory costs and poor robustness.}  
Therefore, performing short-time Fourier transform (STFT), we analyze the signals in T-F domain, in which reverberation can be modeled using a series of cross-band filters~\cite{avargel2007system}.
\addnote[CTF]{1}{Since the energy of cross-band filters is concentrated and more crossband filters may lead to a higher variance in estimation error~\cite{avargel2007system}, we consider only band-to-band filters to simplify the analysis, resulting in CTF approximation~\cite{talmon2009relative}.
Theoretical and experimental error analysis of CTF approximation appears in \cite{avargel2007system} and \cite{tang2025personal}.}
The observation in the T-F domain is
\begin{equation}
    \label{eq:signal_model_stft}
    \begin{aligned}
        X(f,t)&\approx\sum_{l=0}^{L-1}H_l(f)S(f,t-l)+W(f,t) \\
        &=\mathbf{H}(f)\mathbf{S}(f,t)+W(f,t),\\
    \end{aligned}
\end{equation}
where $f$ and $t$ are the indices of frequency band and STFT frame, respectively; $L$ is the length of CTF filter; $X(f,t)$, $S(f,t)$, $W(f,t)$, and $H_l(f)$ are the complex-valued observation signal, source speech signal, noise signal, and CTF coefficient, respectively; 
$\mathbf{H}(f)=\left[H_{L-1}(f),\cdots,H_0(f)\right]\in\mathbb{C}^{1 \times L}$, $\mathbf{S}(f,t)=\left[S(f,t-L+1),\cdots,S(f,t)\right]^T\in \mathbb{C}^{L \times 1}$.


Furthermore, the observation $X(f,t)$, the anechoic source $S(f,t)$, and noise $W(f,t)$ are modeled as random signals.
We have the following assumptions regarding their prior distributions.

\begin{figure}[H]
    \centering
    \includegraphics[width=0.32\textwidth]{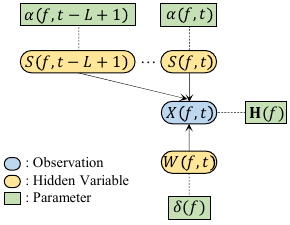}
    \caption{Probabilistic graphical model in VINP.}
    \label{fig:prob_model}
\end{figure}

\begin{itemize}
\item Assumption 1: The anechoic source speech $S(f,t)$ follows a time-variant zero-mean complex-valued Gaussian prior distribution, while the noise signal $W(f,t)$ follows a time-invariant zero-mean complex-valued Gaussian prior distribution. 
Therefore, we have
\begin{equation}
\label{eq:priors}
\left\{
\begin{aligned}
    &S(f,t)\sim\mathcal{CN}\left(0,\alpha^{-1}(f,t)\right)\\
    &W(f,t)\sim\mathcal{CN}\left(0,\delta^{-1}(f)\right),\\
\end{aligned}
\right.
\end{equation}
where $\alpha(f,t)$ and $\delta(f)$ are the precisions of the Gaussian distributions.

\item Assumption~2: 
Both the anechoic source speech signal $S(f,t)$ and the noise signal $W(f,t)$ are statistically independent across all T-F bins.
Defining $\mathbf{S}=[S(1,1),\cdots,S(1,T),\cdots,S(F,T)]^T\in \mathbb{C}^{1\times FT}$ and $\mathbf{W} = [W(1,1),\cdots,W(1,T),\cdots,W(F,T)]^T\in \mathbb{C}^{1\times FT}$, we have
\begin{equation}
\label{eq:priors2}
\left\{
\begin{aligned}
    &\mathbf{S} \sim \prod_{f=1}^F\prod_{t=1}^T p\left(S(f,t)\right)\\
    &\mathbf{W}\sim\prod_{f=1}^F\prod_{t=1}^T p\left(W(f,t)\right).\\
\end{aligned}
\right.
\end{equation}

\item Assumption~3:
The anechoic source speech $\mathbf{S}$ and noise $\mathbf{W}$ are independent of each other, which means
\begin{equation}
    \mathbf{S},\mathbf{W} \sim p(\mathbf{S})p(\mathbf{W}).
\end{equation}

\end{itemize}
With all the aforementioned assumptions, the conditional probability of the observation signal can be expressed as
\begin{equation}
\left\{
\begin{aligned}
    &X(f,t)|\mathbf{S}(f,t);\boldsymbol{\theta}(f)\sim\mathcal{CN}\left(\mathbf{H}(f)\mathbf{S}(f,t),\delta^{-1}(f)\right) \\
    &\mathbf{X}|\mathbf{S};\boldsymbol{\theta}\sim\prod_{f=1}^F\prod_{t=1}^T p\left(X(f,t)|\mathbf{S}(f,t);\boldsymbol{\theta}(f)\right),\\
\end{aligned}
\right.
\end{equation}
where $\boldsymbol{\theta}(f)=\left\{\delta(f),\mathbf{H}(f)\right\}$ and $\boldsymbol{\theta}=\left\{\boldsymbol{\theta}(f)|_{f=1}^{F}\right\}$.
The probabilistic graphical model, which describes the generation process of reverberant recording, is shown in Fig. \ref{fig:prob_model}.


%
%


\subsection{Task Description}

Our goal is to jointly achieve speech dereverberation and blind RIR identification by solving all hidden variables and parameters in Fig. \ref{fig:prob_model} via VBI.

For speech dereverberation, our focus is on the anechoic speech spectrum, which corresponds to the hidden variables in our probabilistic graphical model.
We aim to get its maximum a posteriori (MAP) estimate given the reverberant recording, written as
\begin{equation}
    \hat{\mathbf{S}}=\argmax\limits_{\mathbf{S}}p(\mathbf{S}|\mathbf{X};\boldsymbol{\theta}),
\end{equation}
where the posterior distribution of anechoic source signal can be represented according to Bayes' rule as 
\begin{equation}
    \label{eq:posterior}
    p(\mathbf{S}|\mathbf{X};\boldsymbol{\theta})=\frac{p(\mathbf{S})p(\mathbf{X}|\mathbf{S};\boldsymbol{\theta})}{\int p(\mathbf{S})p(\mathbf{X}|\mathbf{S};\boldsymbol{\theta}) \mathrm{d}\mathbf{S}}.
\end{equation}

For blind RIR identification, our focus is on the CTF filter, which corresponds to the model parameters in our probabilistic model.
Defining the CTF filter in all frequency bands as $\mathbf{H}=\left[\mathbf{H}(1),\cdots,\mathbf{H}(F)\right]$, we aim to obtain its maximum likelihood (ML) estimate, written as
\begin{equation}
    \hat{\mathbf{H}}=\argmax\limits_{\mathbf{H}}p(\mathbf{S},\mathbf{X};\boldsymbol{\theta}).
\end{equation}
The CTF filter provides a representation of RIR in the T-F domain.
We reconstruct the RIR waveform from CTF coefficients through a pseudo measurement process, as detailed in Section \ref{sec:prop}.


\section{Proposed Method}
\label{sec:prop}
\begin{figure}[H]
    \centering
    \includegraphics[width=0.65\linewidth]{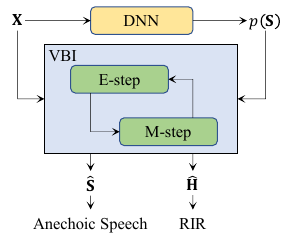}
    \caption{Overview of VINP.}
    \label{fig:overview}
\end{figure}

In VINP, we propose using an arbitrary discriminative dereverberation DNN to estimate the prior distribution of anechoic speech and using VBI to analytically estimate the anechoic spectrum and CTF filter.
Subsequently, we use a pseudo measurement process to transform the CTF filter into the RIR waveform.
The overview of VINP is shown in Fig. \ref{fig:overview}.

\subsection{Estimation of Anechoic Speech Prior}

The direct-path speech signal, which is a scaled and delayed version of the anechoic source speech signal, is free from noise and reverberation.
To avoid estimating the arbitrary direct-path propagation delay and attenuation, instead of the actual source speech, we setup the direct-path speech as the source speech and our target signal, which is still denoted as $s(n)$ or $\mathbf{S}$.  Correspondingly, the RIR begins with the impulse response of the direct-path propagation.

In VINP, we consider the periodogram of the direct-path speech signal as the variance of the anechoic speech prior~$p(\mathbf{S})$.
In each T-F bin, the ideal precision parameter $\alpha(f,t)$ in Eq. (\ref{eq:priors}) is
\begin{equation}
\label{eq:alpha_hat}
    \alpha(f,t) = 1/{|S(f,t)|^{2}}.
\end{equation}
Because Eq. (\ref{eq:alpha_hat}) is unavailable in practice, we propose using an arbitrary discriminative dereverberation DNN to estimate the anechoic speech prior~$p(\mathbf{S})$.
DNN constructs a mapping from reverberant magnitude spectrum $|\mathbf{X}|$ to the anechoic magnitude spectrum $| {\hat{\mathbf{S}}}_{\mathrm{N}}|$ as 
\begin{equation}
\label{eq:fdnn}
    |{\hat{\mathbf{S}}}_{\mathrm{N}}|=f_{\mathrm{DNN}}\left(\left|\mathbf X\right|\right).
\end{equation}
Correspondingly, the estimated anechoic speech prior becomes
\begin{equation}
\label{eq:approxprior}
    p(\hat{\mathbf{S}}_\mathrm{N})=\prod_{f=1}^F\prod_{t=1}^{T}\mathcal{C}\mathcal{N}\left(0,\alpha_{\mathrm{N}}^{-1}(f,t)\right),
\end{equation}
where
\begin{equation}
\label{eq:alpha_hat2}
    \alpha_{\mathrm{N}}(f,t) = 1/{|\hat{S}_{\mathrm{N}}(f,t)|^{2}}.
\end{equation}
Similar to $|S(f,t)|$, $|\hat{S}_{\mathrm{N}}(f,t)|$ is an element in $|{\hat{\mathbf{S}}}_{\mathrm{N}}|$ that corresponds to frequency band $f$ and frame $t$.

Regarding the loss function, we employ the average Kullback-Leibler (KL) divergence~\cite{kullback1951information} to measure the distance of the estimated prior distribution $p(\hat{\mathbf{S}}_\mathrm{N})$ and oracle prior distribution $p(\mathbf{S})$ as
\begin{equation}
\begin{aligned}
&\mathcal{L} = \mathrm{E}_{\mathrm{data}}\left[\frac{\mathrm{KL}\left(p(\hat{\mathbf{S}}_\mathrm{N})||p(\mathbf{S})\right)}{FT}\right]\\
&=\mathrm{E}_{\mathrm{data},f,t}\left[\ln\left({\frac{|S(f,t)|^2+\epsilon}{|\hat{S}_{\mathrm{N}}(f,t)|^2+\epsilon}}\right)+{\frac{|\hat{S}_{\mathrm{N}}(f,t)|^2+\epsilon}{|S(f,t)|^2+\epsilon}}-1\right],\\
\end{aligned}
\end{equation}
where $\epsilon$ is a small constant to avoid numerical instabilities.
\addnote[lossfunc]{1}{
Such a loss function differs from that in conventional discriminative dereverberation DNNs, which directly estimate the anechoic spectrum, yet it shares the same formulation as the Itakura-Saito divergence between spectrograms employed in \cite{bie2022unsupervised,baby2021speech,wang2024rvae,fevotte2009nonnegative}.
In preliminary experiments, we have also explored various loss functions, including mean squared error (MSE) and mean absolute error (MAE) for linear and logarithmic magnitude spectra, alongside backward KL divergence. 
The experimental results confirm that the KL loss outperforms all the alternatives.
}

Following the DNN, approximations $p({\mathbf{S}})\approx p(\hat{\mathbf{S}}_\mathrm{N})$ and $\alpha(f,t)\approx\alpha_{\mathrm{N}}(f,t)$ are employed in the subsequent VBI procedure.
\addnote[dependency]{1}{Notice that the independent assumption of anechoic speech prior is merely intended to simplify the subsequent derivation of VBI.
The DNN learning of speech prior is not limited to the exclusion of correlations between T-F bins.}



\subsection{Variational Bayesian Inference }
Given the estimated prior distribution of anechoic speech and the observation, we solve all hidden variables and parameters via Bayesian inference.

The posterior distribution $p(\mathbf{S}|\mathbf{X};\boldsymbol{\theta})$ is intractable due to the integral term in Eq.~(\ref{eq:posterior}).
Therefore, we turn to VBI, a powerful tool for resolving hierarchical probabilistic models~\cite{bianco2019machine}.
More specifically, we employ the variational expectation-maximization (VEM) algorithm, which provides a way for approximating the complex posterior distribution $p(\mathbf{S}|\mathbf{X};\boldsymbol{\theta})$ with a factored distribution $q(\mathbf{S})$ according to the mean-field theory (MFT) as
\begin{equation}
    p(\mathbf{S}|\mathbf{X};\boldsymbol{\theta}) \approx q(\mathbf{S})=\prod_{f=1}^{F}\prod_{t=1}^{T} q\left(S(f,t)\right).
\end{equation}
After factorization, VEM estimates the posterior $q(\mathbf{S})$ and model parameters $\boldsymbol{\theta}$ by E-step and M-step respectively and iteratively as
\begin{equation}
    \label{eq:estposterior}
    \ln{q\left(S(f,t)\right)}=\left<\ln{p(\mathbf{S},\mathbf{X};\boldsymbol{\theta})}\right>_{\mathbf{S}\backslash S(f,t)}
\end{equation}
and
\begin{equation}
    {\hat{\boldsymbol{\theta}}=\argmax\limits_{\boldsymbol{\theta}}\left<\ln{p(\mathbf{S},\mathbf{X};\boldsymbol{\theta})}\right>_{\mathbf{S}},}
\end{equation}
where $\backslash$ denotes the set subtraction and $\left<\cdot\right>_\cdot$ denotes expectation.
For simplicity, in the subsequent descriptions, we omit the subscript of $\left<\cdot\right>_{\mathbf{S}}$.
We do not update the precision parameter $\alpha(f,t)$ during iterations to prevent the degradation of prior distribution.

The specific formulas are as follows:

\subsubsection{E-step}
In this step, we update the posterior distribution of the anechoic spectrum given the observation and estimated model parameters.
Substitute the probabilistic model into Eq.~(\ref{eq:estposterior}), we have
\begin{equation}
    \begin{aligned}
        &\ln{q\left({S}(f,t)\right)} \\
        &=\left<\ln{p({\mathbf{S}},\mathbf{X};\boldsymbol{\theta})}\right>_{{\mathbf{S}}\backslash {S}(f,t)}\\
        &=
        \ln{p\left({S}(f,t)\right)}\\
        &\quad+\left<\sum_{l=0}^{L-1}\ln{p\left(X(f,t+l)|{\mathbf{S}}(f,t+l);\boldsymbol{\theta}(f)\right)}\right>_{{\mathbf{S}}\backslash {S}(f,t)}+c,
    \end{aligned}
\end{equation}
where $c$ is a constant term that is independent of ${S}(f,t)$, and
\begin{equation}
\left\{
\begin{aligned}
    &\ln{p\left({S}(f,t)\right)}=\ln{\alpha(f,t)}-\alpha(f,t)\left|{S}(f,t)\right|^2+c \\
    &\ln{p\left(X(f,t+l)|{\mathbf{S}}(f,t+l);\boldsymbol{\theta}(f)\right)} \\
    &\quad = \ln{\delta(f)}-\delta(f)\left|X(f,t+l)-\mathbf{H}(f){\mathbf{S}}(f,t+l)\right|^2+c.\\ 
\end{aligned}
\right.
\end{equation}

According to the property of Gaussian distribution, the posterior distribution is also Gaussian, written as 
\begin{equation}
    q\left({S}(f,t)\right)=\mathcal{CN}\left({\mu}(f,t),{\gamma}^{-1}(f,t)\right),
\end{equation}
whose precision and mean have closed-form 
solutions as
\begin{equation}
\label{eq:mu_var}
\left\{
\begin{aligned}
    &\hat{\gamma}(f,t)=\alpha(f,t)+\delta(f) ||\mathbf{H}(f)||_2^2 \\
    &\hat{\mu}(f,t)={\gamma}^{-1}(f,t)\delta(f)\\
    &\quad\times\left[\sum_{l=0}^{L-1}H_l^*(f)\left[X(f,t+l)-\mathbf{H}_{\backslash l}(f)\hat{\boldsymbol{\mu}}_{\mathrm{pre}}(f,t+l)\right]\right],
\end{aligned}
\right.
\end{equation}
where $\mathbf{H}_{\backslash l}(f)$ is same as $\mathbf{H}(f)$ except that $H_l(f)$ is set to $0$, 
and $\hat{\boldsymbol{\mu}}_{\mathrm{pre}}(f,t+l)=\left[{\hat\mu}_{\mathrm{pre}}(f,t+l-L+1),\cdots,{\hat\mu}_{\mathrm{pre}}(f,t+l)\right]^T$ contains the estimates of means from the previous VEM iteration.

In order to make VEM converge more smoothly, we further apply an exponential moving average (EMA) to the estimates in Eq.~(\ref{eq:mu_var}) as
\begin{equation}
\label{eq:mu_var_sm}
\left\{
\begin{aligned}
    &\hat{\gamma}^{-1}(f,t)=\lambda\hat\gamma^{-1}_{\mathrm{pre}}(f,t)+(1-\lambda)\hat\gamma^{-1}(f,t)\\
    &\hat{\mu}(f,t)=\lambda\hat{\mu}_{\mathrm{pre}}(f,t)+(1-\lambda){\hat\mu}(f,t),
\end{aligned}
\right.
\end{equation}
where $\lambda$ is a smoothing factor, $\hat{\mu}_{\mathrm{pre}}(f,t)$ and $\hat\gamma^{-1}_{\mathrm{pre}}(f,t)$ are the estimates from the previous VEM iteration. 
The mean of the posterior distribution is the MAP estimate of anechoic spectrum $\hat{\mathbf{S}}$.
It is also worth noticing that the E-step can be implemented in parallel across all T-F bins.

\subsubsection{M-step}
In this step, VEM updates the noise precision and CTF filter by maximizing the expected logarithmic likelihood of the complete data, which is
\begin{equation}
    \begin{aligned}
        &\left<\ln{p\left({\mathbf{S}},\mathbf{X};\boldsymbol{\theta}\right)}\right>\\
        &=
        \left<\ln{p\left(\mathbf{X}|{\mathbf{S}};\boldsymbol{\theta}\right)}\right> +c\\
        &= T\ln{\delta(f)}-\delta(f)\left<\sum_{t=1}^{T}\left|X(f,t)-\mathbf{H}(f){\mathbf{S}}(f,t)\right|^2\right> +c, \\
    \end{aligned}
\end{equation}
where $c$ is a constant term that is independent of $\delta(f)$ and $\mathbf{H}(f)$.
Setting the first derivative with respect to the parameters to zero, the noise precision is updated as
\begin{equation}
\label{eq:est_delta}
\begin{aligned}
    &\hat{\delta}(f)\\
    &= T\bigg/\sum\limits_{t=1}^T\left[\left|X(f,t)\right|^2-2\mathrm{Re}\left\{X^*(f,t)\mathbf{H}(f)\left<{\mathbf{S}}(f,t)\right>\right\}\right.\\
    &\quad +\left.{\mathbf{H}(f)\left<{\mathbf{S}}(f,t){\mathbf{S}}^H(f,t)\right>\mathbf{H}^H(f)}\right],
\end{aligned}
\end{equation}
and the CTF filter is updated as
\begin{equation}
\label{eq:est_H}
    \begin{aligned}
        &\hat{\mathbf{H}}(f) \\
        &= \left[\sum_{t=1}^{T}
        X(f,t)\left<{\mathbf{S}}^H(f,t)\right>\right]
        \left[\sum_{t=1}^{T}\left<{\mathbf{S}}(f,t){\mathbf{S}}^H(f,t)\right>\right]^{-1},
    \end{aligned}
\end{equation}
where
\begin{equation}
    \left<\mathbf{S}(f,t)\right>=\boldsymbol{\mu}(f,t)=\left[\mu(f,t-L+1),\cdots,\mu(f,t)\right]^T,
\end{equation}
and
\begin{equation}
\begin{aligned}
    &\left<\mathbf{S}(f,t)\mathbf{S}^H(f,t)\right>\\
    &=\boldsymbol{\mu}(f,t)\boldsymbol{\mu}^H(f,t)\\
    &\quad +\mathrm{diag}([\gamma^{-1}(f,t-L+1),\cdots,\gamma^{-1}(f,t)]).
\end{aligned}
\end{equation}
$\mathrm{diag}(\cdot)$ denotes the operation of constructing a diagonal matrix.
Just like the E-step, the M-step can also be implemented in parallel across all T-F bins.

\subsubsection{Initialization of VEM Parameters}

The initialization of parameters plays a crucial role in the convergence of VEM.
We use an uninformative initialization in VINP.
Before iteration, the mean and variance of the anechoic speech posterior $p(\mathbf{S})$ are set 
to zero and the spectrogram of reverberant recording respectively as
\begin{equation}
\label{eq:init1}
\left\{
\begin{aligned}
    &\gamma(f,t)=1/\left|X(f,t)\right|^{2}\\
    &\mu(f,t)=0.\\
\end{aligned}
\right.
\end{equation}
The CTF coefficients in each frequency band are set to 0, except that the first coefficient is set to 1, which means
\begin{equation}
\label{eq:init2}
H_{l}(f)=
\left\{
\begin{aligned}
    &1, \quad l=0\\
    &0, \quad 1\leq l \leq L-1.\\
\end{aligned}
\right.
\end{equation}
Because even during speech activity, the short-term power spectral density of observation often decays to values representative of the noise power level~\cite{martin2001noise}, the initial variance of the additional noise in each frequency band is set to the minimum power across all frames, which means
\begin{equation}
\label{eq:init3}
    \delta(f) = \left[\min\limits_t\left(\left|X(f,t)\right|^2\right)\right]^{-1}.
\end{equation}

\subsubsection{\addnote[]{1}{Outputs of VBI}}
\addnote[stoprule]{1}{
During VEM iteration, in each frequency band, there is a narrow-band expected logarithmic likelihood of complete data, written as $\left<\ln p\left(\mathbf{S}(f),\mathbf{X}(f);\boldsymbol{\theta}(f)\right)\right>$.
Once VEM reaches the preset maximum number of iterations, it terminates.
In each frequency band, the output of VBI procedure is the estimated complex-valued clean spectrum $\hat{\mathbf{S}}(f)$ and CTF filter $\hat{\mathbf{H}}(f)$ corresponding to the maximum expected logarithmic likelihood.}

The VBI procedure is summarized in Algorithm~\ref{algo:VBI}.
The prior distribution of anechoic speech serves to constrain the infinite set of potential solutions inherent in the signal model, which is the key to effective dereverberation~\cite{10638210}.
Note that in the proposed framework, the prior distribution of anechoic speech remains unchanged during iterations.
Therefore, the DNN performs inference through a single forward pass without iterative optimization.

\begin{algorithm}[H]
    \renewcommand{\algorithmicrequire}{\textbf{Input:}}
	\renewcommand{\algorithmicensure}{\textbf{Output:}}
	\caption{VBI procedure.}
    \label{algo:VBI}
    \begin{algorithmic}[1] 
        \REQUIRE Reverberant microphone recording $\mathbf{X}$; 
	    \ENSURE Anechoic speech spectrum estimate $\hat{\mathbf{S}}$ and CTF filter estimate $\hat{\mathbf{H}}$; 

        \STATE Estimate prior distribution of anechoic speech using Eq. (\ref{eq:fdnn}), Eq. (\ref{eq:approxprior}) and Eq. (\ref{eq:alpha_hat2});
        \STATE Initialize VEM parameters using Eq. (\ref{eq:init1}), Eq. (\ref{eq:init2}) and Eq. (\ref{eq:init3});
        \REPEAT
            \STATE E-step: update the posterior distribution of anechoic speech using Eq. (\ref{eq:mu_var}) and Eq. (\ref{eq:mu_var_sm});
            \STATE M-step: update the parameter estimates of the signal model using Eq. (\ref{eq:est_delta}) and Eq. (\ref{eq:est_H});
        \UNTIL{Reach the maximum number of iterations.}
    \end{algorithmic}
\end{algorithm}

\subsection{Transformation to Waveforms}
After the VBI procedure, both the anechoic spectrum and the CTF filter are estimated.
We need to further transform these T-F representations into waveforms.
By applying inverse STFT to the anechoic spectrum, we can easily obtain the anechoic speech waveform.
However, CTF is the model of reverberation in the T-F domain rather than the STFT of RIR.
Therefore, we cannot convert the CTF filter to RIR directly.
In VINP, we design a pseudo intrusive measurement process to obtain the RIR estimate as follows.

Reviewing existing approaches, a common method for intrusive measuring the RIR of an acoustical system is to apply a known excitation signal and measure the microphone recording~\cite{stan2002comparison}.
When a loudspeaker plays an excitation signal $e(n)$, the noiseless microphone recording $y(n)$ can be written as
\begin{equation}
    y(n) = h(n)*e(n),
\end{equation}
where $h(n)$ has the same meaning as in Eq.~(\ref{eq:signal_model_time}).
A commonly used logarithmic sine sweep excitation signal can be expressed as~\cite{stan2002comparison,farina2000simultaneous}
\begin{equation}
\label{eq:excitation}
    e(n)=\sin \left[\frac{N\omega_1}{\ln \left({\omega_2}/{\omega_1}\right)}\left(e^{n\ln \left({\omega_2}/{\omega_1}\right)/N}-1\right)\right],
\end{equation}
where $\omega_1$ is the initial radian frequency and $\omega_2$ is the final radian frequency of the sweep with duration $N$.
Through an ideal inverse filter $v(n)$, the excitation signal can be transformed into an impulse $\delta(n)$, as
\begin{equation}
    e(n)*v(n)=\delta(n).
\end{equation}
For the logarithmic sine sweep excitation, the inverse filter $v(n)$ is an amplitude-modulated and time-reversed version of itself~\cite{stan2002comparison,farina2000simultaneous}.
Subsequently, the RIR can be estimated by convolving the measurement $y(n)$ with the inverse filter $v(n)$ as
\begin{equation}
    h(n)=y(n)*v(n).
\end{equation}

In our approach, the excitation signal is convoluted with the CTF filter in the T-F domain along the time dimension to yield a pseudo measurement spectrum $\tilde{Y}(f,t)$ as
\begin{equation}
\label{eq:pseudo_measure}
\tilde{Y}(f,t)={\mathbf{H}}(f)\mathbf{E}(f,t),
\end{equation}
where $\mathbf{E}(f,t)=\left[E(f,t-L+1),\cdots,E(f,t)\right]^T\in \mathbb{C}^{L \times 1}$, $\tilde{Y}(f,t)$ and $E(f,t)$ are the STFT coefficient of $\tilde{y}(n)$ and $e(n)$, respectively.
Applying inverse STFT to $\tilde{Y}(f,t)$, we use the pseudo measurement $\tilde{y}(n)$ and the inverse filter $v(n)$ to estimate the RIR waveform as
\begin{equation}
\label{eq:inverse_filtering}
    \hat{h}(n)=\tilde{y}(n)*v(n).
\end{equation}
The transformation from the CTF filter to RIR waveform is summarized in Algorithm~\ref{algo:RIR}.
\begin{algorithm}[H]
    \renewcommand{\algorithmicrequire}{\textbf{Input:}}
	\renewcommand{\algorithmicensure}{\textbf{Output:}}
	\caption{Transformation from CTF to RIR.}
    \label{algo:RIR}
    \begin{algorithmic}[1] 
        \REQUIRE CTF filter ${\mathbf{H}}$; 
	    \ENSURE RIR waveform estimate $\hat{h}(n)$; 
        \STATE Pick a pair of excitation signal $e(n)$ and its inverse filter $v(n)$;
        \STATE Build a pseudo measurement signal $\tilde{y}(n)$ using Eq. (\ref{eq:pseudo_measure});
        \STATE Estimate RIR $\hat{h}(n)$ by inverse filtering as Eq. (\ref{eq:inverse_filtering}).

    \end{algorithmic}
\end{algorithm}


\section{Experiments}
\label{sec:expset}

\subsection{Datasets}
VINP is designed for joint speech dereverberation and blind RIR identification.
We use a single training set and two different test sets.
\addnote[]{1}{To ensure sampling rate consistency, all audio utterances in these datasets are downsampled to 16 kHz.}
\subsubsection{{Training Set}}
The training utterances are generated by convolving anechoic source speech with reverberant and direct-path RIRs, followed by noise addition.
The source speech consists of 200 hours of high-quality English speech utterances, including the clean speech from DNS Challenge~\cite{reddy2020interspeech} and VCTK~\cite{valentini2016investigating} that have a raw DNSMOS p.835 score~\cite{reddy2022dnsmos} exceeding 3.5, as well as the whole EARS~\cite{richter2024ears} corpus.
We simulate 100,000 pairs of reverberant and direct-path RIRs using the gpuRIR toolbox~\cite{diaz2021gpurir}.
The simulated speaker and omnidirectional microphone are randomly placed in rooms with dimensions randomly selected within a range of 3 m to 15 m for length and width, and 2.5 m to 6 m for height.
The minimum distance between the speaker/microphone and the wall is 1 m. 
Reverberant RIRs have RT60s uniformly distributed within the range of 0.2 s to 1.5 s. 
Direct-path RIRs are generated using the same geometric parameters as the reverberant ones but with an absorption coefficient of 0.99.
Noise recordings from NOISEX-92~\cite{varga1993assessment} and the training set of REVERB Challenge~\cite{kinoshita2013reverb} are used.
The signal-to-noise ratio (SNR) is uniformly distributed within the range of 5 dB to 20 dB.


\subsubsection{{Test Set for Speech Dereverberation}}
For speech dereverberation, we utilize the official single-channel test set from the REVERB Challenge~\cite{kinoshita2013reverb}, which includes both simulated recordings (marked as 'SimData') and real recordings (marked as 'RealData').

In SimData, there exist six distinct reverberation conditions: three room volumes (small, medium, and large), and two distances between the speaker and the microphone (50 cm and 200 cm). The RT60 values are approximately 0.25 s, 0.5 s, and 0.7 s. 
The noise is stationary, mainly generated by air conditioning systems. 
SimData has a SNR of 20 dB.

RealData consists of utterances spoken by human speakers in a noisy and reverberant meeting room. It includes two reverberation conditions: one room and two distances between the speaker and the microphone array (approximately 100 cm and 250 cm). The RT60 is about 0.7 s.

\subsubsection{{Test Set for Blind RIR Identification}}
A test set named 'SimACE' is constructed to evaluate blind RIR identification.
In SimACE, microphone signals are simulated by convolving the clean speech from the 'si\_et\_05' subset in WSJ0 corpus~\cite{paul1992design} with the downsampled recorded RIRs from the 'Single' subset in ACE Challenge~\cite{eaton2016estimation}, and adding noise from the test set in REVERB Challenge~\cite{kinoshita2013reverb}.
The minimum and maximum RT60s are 0.332~s and 1.22 s, respectively.
More details about the RIRs can be found in~\cite{eaton2016estimation}.
SimACE has a SNR of 20 dB.

\addnote[]{1}{Notice that, for both tasks, the RIRs in the test sets are real-measured instead of simulated, which may lead to some mismatch with the training set.}

\subsection{Implementation of VINP}
\subsubsection{Data Representation}
Before feeding the speech into VINP, the reverberant waveform is normalized by its maximum absolute value. 
The STFT analysis window and synthesis window are set to Hann windows with a length of 512 samples and 75\% overlap.

\begin{figure}[H]
    \centering
    \subfloat[VINP-TCN+SA+S]{
    \label{fig:DNNarch1}
        \centering
        {\includegraphics[width=0.35\linewidth]{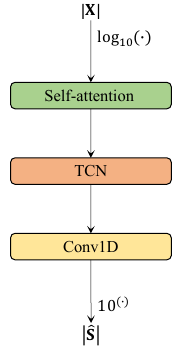}}
    }
    \subfloat[VINP-oSpatialNet]{
    \label{fig:DNNarch2}
        \centering
        {\includegraphics[width=0.35\linewidth]{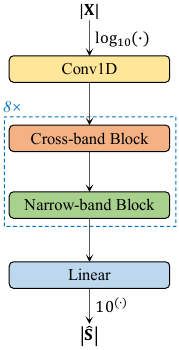}}
    }\hfill
    \caption{DNN architectures in VINP.}
    \label{fig:DNNarch}
\end{figure}

\subsubsection{DNN Architecture}

VINP is able to employ any discriminative dereverberation DNNs to estimate the prior distribution of anechoic speech.
\addnote[]{1}{
Two representative dereverberation backbones TCN+SA+S~\cite{zhao2020monaural} and oSpatialNet-Mamba~\cite{10570301} are selected.
Among those backbones, TCN+SA+S includes a self-attention module to produce dynamic representations given input features, a temporal convolutional network (TCN) to learn a nonlinear mapping from such representations to the magnitude spectrum of anechoic speech, and a 1-D convolution module to smooth the enhanced magnitude among adjacent frames.
Different from TCN+SA+S, oSpatialNet-Mamba employs a 1-D convolution input layer and a linear output layer, and the so-called stacked cross-band and narrow-band blocks to build a mapping between complex-valued spectrums.
The narrow-band block, composed of Mamba layers and convolutional layers along the time axis, processes each frequency independently with the same network parameters.
The cross-band block, composed of convolutional layers along the frequency axis and full-band linear modules, processes each frame independently with the same network parameters.
oSpatialNet-Mamba was initially proposed for online multi-channel applications. 
However, for single-channel applications, the spatial information still lies in RIR.
The narrow-band block, designed based on CTF approximation, is able to learn such spatial information and thereby enabling single-channel speech dereverberation.
Subsequent research has evaluated the effectiveness of oSpatialNet for single-channel speech enhancement ~\cite{11097896}.}

\addnote[selectbackbone]{1}{
In VINP, we adopt TCN+SA+S and a modified version oSpatialNet-Mamba, resulting in two different versions marked as \textbf{'VINP-TCN+SA+S'} and \textbf{'VINP-oSpatialNet'}, respectively.
For both versions, the input and output of DNNs are the 10-based logarithmic magnitude spectra.
To avoid numerical issues, a small constant $10^{-8}$ is added to the magnitude spectra before taking the logarithm.
In VINP-TCN+SA+S, we use the same structure and settings as the original TCN+SA+S as shown in Fig. \ref{fig:DNNarch1}, except that there is no activation function after the output layer, and we do not use dropout.
In VINP-oSpatialNet, the DNN utilizes a 1-D convolution layer with a kernel size of $3$ to expand the input to 96 dimensions.
The 'cross-band block' and 'narrow-band block' follow the original definition in oSpatialNet-Mamba, except that we replace the second forward Mamba layer in the narrow-band block with a backward Mamba layer by simply reversing the input and output along the time dimension for offline processing.
The remaining modules are the same as those in the original oSpatialNet-Mamba, as depicted in Fig. \ref{fig:DNNarch2}.
We use VINP-TCN+SA+S and VINP-oSpatialNet to demonstrate the performance of two DNNs with different architectures and capabilities.}

\subsubsection{Training Configuration}
For DNN training, we set the hyperparameter in the loss function to $\epsilon=0.0001$.
The speech utterances are segmented into 3 s.
\addnote[]{1}{Each epoch contains 97,092 samples.}
The batch sizes of VINP+TCN+SA+S and VINP-oSpatialNet are set to 16 and 4, respectively.
The AdamW optimizer~\cite{loshchilov2017decoupled} with an initial learning rate of 0.001 is employed.
The learning rate exponentially decays with $\mathrm{lr} \leftarrow{} 0.001\times0.97^{\mathrm{epoch}}$ and $\mathrm{lr} \leftarrow{} 0.001\times0.9^{\mathrm{epoch}}$ in VINP-TCN+SA+S and VINP-oSpatialNet, respectively.
Gradient clipping is applied with a L2-norm threshold of 10.
The training is carried out for 800,000 steps in total.
We average the model weights of the best three epochs as the final model.

\subsubsection{VBI Settings}
The length of the CTF filter is set to $L=30$. The fixed smoothing factor $\lambda$ in Eq. (\ref{eq:mu_var_sm}) is set to 0.7 to obtain a stable result.
Since the fundamental frequency of human speech is always higher than 85 Hz~\cite{howard2007voice}, we ignore the three lowest frequency bands and set their coefficients to zero to avoid the effect of extremely low SNR in these bands. 
Therefore, VBI processes a total of 254 frequency bands.

\subsubsection{Pseudo Excitation Signal}
We use a logarithmic sine sweep signal with a frequency range of 62.5 Hz to 8000 Hz and a duration of 8.192 s as the pseudo excitation signal.
Fade-in and fade-out with 256 and 128 samples are applied to the excitation signal to mitigate spectral leakage.
The formulas for the excitation signal and the corresponding inverse filter can be found in Eq. (\ref{eq:excitation}) and \cite{stan2002comparison,farina2000simultaneous}.

\subsubsection{RT60 and DRR Estimation}
RT60 and DRR are key acoustic parameters characterizing the properties of RIR, and thus serve to evaluate the accuracy of RIR identification.

RT60 is the time required for the sound energy in an enclosure to decay by 60 dB after the sound source stops.
Given the RIR waveform, Schroeder’s integrated energy decay curve (EDC) is calculated as~\cite{schroeder1965new}
\begin{equation}
\label{eq:edc}
    \mathrm{EDC}(n)=\sum_{m=n}^{\infty}h^2(m).
\end{equation}
Since sound energy decays exponentially over time, a linear fitting is applied to a segment of the logarithmic EDC, and the slope is used to calculate RT60. 
In this process, the key to linear fitting lies in the heuristic strategy for selecting the fitting range. 
In this work, the starting sampling point of the fitting is chosen within the range from 5 dB below the direct-path peak to the point corresponding to a 50 ms delay after the direct path. 
Meanwhile, the ending sampling point is defined as a point with 5 dB attenuation relative to the starting point. 
We fit all intervals that meet the conditions and adopt the result with the maximum absolute Pearson correlation coefficient for the RT60 estimation process as
\begin{equation}
\label{eq:t60}
    T_{60}=-60/k,
\end{equation}
where $k$ is the slope of the fitted line.

DRR refers to the ratio of the direct-path sound energy to the reverberant sound energy.
In this work, the DRR is defined as
\begin{equation}
\label{eq:drr}
    \mathrm{DRR}=10\log_{10}\frac{\sum_{n=n_d-\Delta n_d}^{n_d+\Delta n_d}h^2(n)}{\sum_{n=0}^{n_d-\Delta n_d}h^2(n)+\sum_{n=n_d+\Delta n_d}^{\infty}h^2(n)},
\end{equation}
where the direct-path signal arrives at the $n_d$th sample, and $\Delta n_d$ is the additional sample spread for the direct-path response, which typically corresponds to a 2.5 ms duration~\cite{eaton2016estimation}.

\subsection{Comparison Methods}
\subsubsection{{Speech Dereverberation}}
We compare VINP with various advanced dereverberation approaches, including GWPE~\cite{yoshioka2012generalization},
SkipConvNet~\cite{kothapally2020skipconvnet},
CMGAN~\cite{abdulatif2024cmgan},
and StoRM~\cite{lemercier2023storm}.
Additionally, comparisons are also made with the backbones TCN+SA+S~\cite{zhao2020monaural} and the modified oSpatialNet-Mamba~\cite{10570301}, which are trained using their original loss functions and input/output formats.
In TCN+SA+S, we use the recommended Griffin-Lim's iterative algorithm~\cite{griffin1984signal} to restore the phase spectrum. 
In the modified oSpatialNet-Mamba,
we use the same DNN as in VINP-oSpatialNet and expand the number of channels in the input and output layers to 2 to process complex-valued spectra.
We mark the modified DNN architecture as 'oSpatialNet*'.
\addnote[]{1}{To further illustrate the impact of VEM iterations, we also show results of VINP-TCN+SA+S and VINP-oSpatialNet without VEM iterations (i.e., using only the DNNs in VINP to enhance the magnitude spectrum and keeping the phase spectrum reverberant, resulting in a spectrum as $|{\hat{\mathbf{S}}}_{\mathrm{N}}|\exp{(j \angle \mathbf{X)}}$). }
All comparison methods are trained and tested with their official codes (if available).
All DL-based approaches are trained from scratch on the same training set.
GWPE is implemented using the NaraWPE python package~\cite{Drude2018NaraWPE}.

\begin{table}[b]
\centering
\renewcommand\arraystretch{1.2}
\caption{Number of parameters and MACs per second\\ for DL-based methods}
\label{tab:para_mac}
\begin{tabular}{c|c|c}
    \Xhline{1pt}
    Method&Params~(M)&MACs~(G/s)\\
    \Xhline{0.4pt}
    SkipConvNet~\cite{kothapally2020skipconvnet}&64.3&11\\
    CMGAN~\cite{abdulatif2024cmgan} & 1.8 & 31\\
    StoRM~\cite{lemercier2023storm} & 27.3+27.8=55.1 & 2300\\
    \cdashline{1-3}
    TCN+SA+S~\cite{zhao2020monaural}& 4.7 & 0.7\\
    oSpatialNet*~\cite{10570301}&1.7&36.6\\
    \Xhline{0.4pt}
    \textbf{VINP-TCN+SA+S (prop.)}& 4.7 & 0.7+0.27$\times$iterations\\
    $\hookrightarrow$ w/o VEM& 4.7 & 0.7\\
    \textbf{VINP-oSpatialNet (prop.)}&1.7&36.6+0.27$\times$iterations\\
    $\hookrightarrow$ w/o VEM&1.7&36.6\\
    \Xhline{1pt}
\end{tabular}
\end{table}
\begin{figure}
    \centering
    \includegraphics[width=0.46\linewidth]{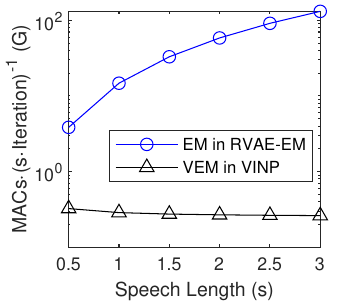}
    \caption{MACs per second and per iteration versus speech length.}
    \label{fig:macs}
\end{figure}
The number of parameters and the multiply-accumulate operations per second (MACs, G/s) of the DL-based approaches are shown in Table \ref{tab:para_mac}.
Additionally, under the same settings of STFT and CTF length, a comparison of MACs per second and per iteration between the VEM algorithm in VINP and the EM algorithm in our previous work RVAE-EM~\cite{wang2024rvae} with regard to speech length is presented in Fig. \ref{fig:macs}.
The asymptotic complexity of the VBI procedure in VINP is $O(FTL^2)$, indicating a linear growth with respect to the speech length, and its MACs is approximately a constant value of 0.27 G/s. 
In contrast, the asymptotic complexity of the EM algorithm in RVAE-EM is $O(FT^3)$ and there is always $T \gg L$.

\subsubsection{{Blind RIR Identification}}
We evaluate the effectiveness of RIR identification by measuring RT60 and DRR from the RIR estimates. 
The comparison methods include Jeub's blind DRR estimation approach~\cite{jeub2011blind}, FiNS~\cite{steinmetz2021filtered} and BUDDy~\cite{lemercier2024unsupervised}.
All these methods are implemented using the official codes (if available). 
Specifically, 
we modified the convolutional kernel size and STFT settings in FiNS to enable the network to process audio with a 16kHz sampling rate.
As an unsupervised approach, we directly use the pretrained BUDDy model without retraining.
For both FiNS and BUDDy, we employ the same implementation as in VINP to estimate RT60s and DRRs, as specified in Eq. (\ref{eq:t60}) and Eq. (\ref{eq:drr}).

\begin{table*}[htbp]
\centering
\renewcommand\arraystretch{1.2}
\caption{\addnote[]{1}{Dereverberation results on REVERB~(1-ch)}}
\label{tab:results_SD}
\resizebox{\textwidth}{!}{
\begin{tabular}{c|c|c|c|c|c|c|c|c|c|c|c|c|c|c}
    \Xhline{1pt}
    \multirow{3}{*}{Method}&\multicolumn{9}{c|}{SimData}&\multicolumn{5}{c}{RealData} \\
    \Xcline{2-15}{0.4pt}
    &\multirow{2}{*}{MOS$\uparrow$}&\multirow{2}{*}{PESQ$\uparrow$}&\multirow{2}{*}{ESTOI$\uparrow$}&\multirow{2}{*}{LSD$\downarrow$}&\multicolumn{2}{c|}{DNSMOS$\uparrow$}&\multicolumn{3}{c|}{WER~(\%)$\downarrow$}&\multicolumn{2}{c|}{DNSMOS$\uparrow$}&\multicolumn{3}{c}{WER~(\%)$\downarrow$}\\
    \Xcline{6-15}{0.4pt}
    &&&&&P.835&P.808&tiny&small&medium&P.835&P.808&tiny&small&medium\\
    \Xhline{0.4pt}
    Unprocessed&2.13&1.48&0.70&5.55&2.26&3.20&13.3&5.7&4.6&1.38&2.82&20.4&7.5&5.6\\
    \textcolor{gray}{Clean}&\textcolor{gray}{4.32}&\textcolor{gray}{-}&\textcolor{gray}{-}&\textcolor{gray}{-}&\textcolor{gray}{3.28}&\textcolor{gray}{3.90}&\textcolor{gray}{7.4}&\textcolor{gray}{4.5}&\textcolor{gray}{4.0}&\textcolor{gray}{-}&\textcolor{gray}{-}&\textcolor{gray}{-}&\textcolor{gray}{-}&\textcolor{gray}{-}\\
    \Xhline{0.4pt}
    GWPE~\cite{yoshioka2012generalization}&-&1.55&0.72&5.42&2.30&3.22&12.4&5.6&4.6&1.47&2.83&18.4&6.9&5.5\\
    SkipConvNet~\cite{kothapally2020skipconvnet}&-&2.12&0.81&3.19&2.91&3.60&13.3&\cellcolor{graybackground}6.3&\cellcolor{graybackground}5.2&2.66&3.40&\cellcolor{graybackground}22.2&\cellcolor{graybackground}9.5&\cellcolor{graybackground}7.3\\
    CMGAN~\cite{abdulatif2024cmgan}&3.27&2.80&0.90&3.86&\textbf{3.34}&3.79&9.6&5.2&4.4&\textbf{3.36}&\textbf{3.99}&10.6&5.5&4.6\\
    StoRM~\cite{lemercier2023storm}&3.97&2.36&0.86&3.59&3.24&\textbf{3.93}&11.0&\cellcolor{graybackground}6.1&\cellcolor{graybackground}5.1&3.20&3.94&17.0&\cellcolor{graybackground}9.2&\cellcolor{graybackground}7.2 \\
    \cdashline{1-14}
    TCN+SA+S~\cite{zhao2020monaural}&2.97&2.60&0.86&3.70&3.12&3.73&12.2&\cellcolor{graybackground}6.6&\cellcolor{graybackground}5.5&3.03&3.73&\cellcolor{graybackground}21.6&\cellcolor{graybackground}12.8&\cellcolor{graybackground}10.2\\
    oSpatialNet*~\cite{10570301}&3.71&\textbf{2.87}&\textbf{0.92}&\textbf{3.03}&3.16&3.88&9.1&5.0&\textbf{4.2}&3.10&3.87&10.5&5.5&4.6\\
    \Xhline{0.4pt}
    VINP-TCN+SA+S (prop.)&4.01&2.52&0.87&3.25&3.10&3.88&9.0&5.1&4.4&2.89&3.77&11.8&6.1&5.2\\
    $\hookrightarrow$ w/o VEM&-&2.27&0.84&3.59&3.09&3.67&11.7&\cellcolor{graybackground}6.2&\cellcolor{graybackground}5.2&2.86&3.37&17.0&\cellcolor{graybackground}8.4&\cellcolor{graybackground}7.1\\
    VINP-oSpatialNet (prop.)&\textbf{4.17}&2.82&0.90&3.17&3.11&3.86&\textbf{8.2}&\textbf{4.9}&\textbf{4.2}&3.00&3.80&\textbf{9.1}&\textbf{5.1}&\textbf{4.4}\\
    $\hookrightarrow$ w/o VEM&-&2.69&0.90&3.33&2.88&3.74&8.3&5.0&\textbf{4.2}&2.98&3.58&9.6&5.2&4.5\\
    \Xhline{1pt}
\end{tabular}}
\end{table*}

\subsection{Evaluation Metrics}
\subsubsection{{Speech Dereverberation}}

Speech dereverberation is evaluated in terms of both perception quality and ASR accuracy. 

We use the commonly used speech quality metrics including perceptual evaluation of speech quality (PESQ)~\cite{rix2001perceptual}, extended short-time objective intelligibility (ESTOI)~\cite{jensen2016algorithm}, and
the overall score of deep noise suppression mean opinion score (DNSMOS) P.808~\cite{reddy2021dnsmos} and P.835~\cite{reddy2022dnsmos}.
Because DNSMOS P.835 is scale-variant, speech utterances are normalized by their maximum absolute value before evaluation. 
Higher scores indicate better speech quality and intelligibility.
\addnote[]{1}{Moreover, log-spectral distortion (LSD)~\cite{du2008speech} in dB is employed to quantify the difference between the enhanced spectrogram and the clean one.
Before calculating LSD, the clean speech is normalized by its maximum absolute value, while the enhanced speech is scaled to match the power of the clean speech.
Lower LSD indicates a better estimation of the spectrogram.}
\addnote[]{1}{
However, while these metrics are widely adopted to evaluate speech quality, none of them serves as the gold standard. 
Therefore, we design a subjective absolute category rating (ACR) listening test on REVERB SimData according to ITU-T Rec. P.808~\cite{P.808}.
In the ACR test, a total of 20 subjects (7 females and 13 males) aged from 20 to 55 years were asked to give overall speech quality scores using five-point category-judgement scales, in which a score of 5 indicates excellent speech quality and 1 indicates poor speech quality.
In the listening test, all methods were applied to process the same 10 inputs, which were randomly selected from the noisy reverberant speech in REVERB SimData. 
The unprocessed and clean utterances were also included to screen the subjects.
During the test, the listeners were blind to the methods, and the speech samples were arranged in a random order.
MOS is used as the final subjective metric. 
Two subjects (1 female and 1 male) demonstrating a consistent preference for unprocessed over clean speech were excluded from further analysis.}

We utilize the popular pre-trained Whisper~\cite{radford2023whisper} ‘tiny' model (with 39 M parameters), 'small' model (with 244 M parameters), and ‘medium’ model (with 769 M parameters) for ASR evaluation. 
No additional dataset-specific finetuning or retraining is applied before ASR inference. 
WER is used as the evaluation metric.
Lower WER indicates better ASR performance.
Since Whisper is scale-variant, all utterances are normalized by their maximum absolute value before being processed.

\subsubsection{{Blind RIR Identification}}
We use the accuracy of RT60 and DRR estimation to evaluate blind RIR identification.
We present MAE and the root mean square error (RMSE) of RT60 and DRR between their estimates and ground-truth values over the SimACE test set.
Lower MAE and RMSE indicate better results.
\addnote[]{1}{Notice that the reference RT60s and DRRs are calculated using Eq. (\ref{eq:t60}) and Eq. (\ref{eq:drr}) to ensure fair comparison.}

\begin{figure*}[ht]
    \centering
    \subfloat[Logarithmic likelihood]{
        \includegraphics[width=0.185\linewidth]{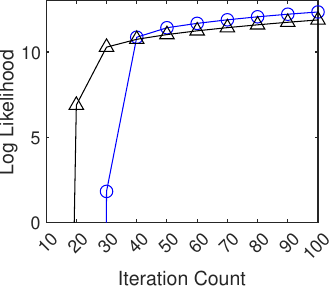}
        }
    \subfloat[PESQ]{
        \includegraphics[width=0.185\linewidth]{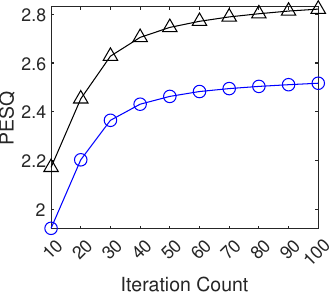}
        }
    \subfloat[ESTOI]{
        \includegraphics[width=0.185\linewidth]{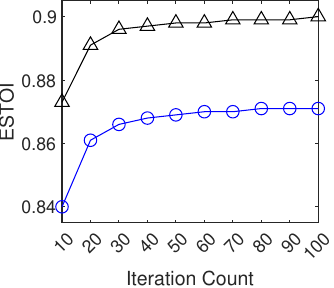}
        }
    \subfloat[DNSMOS]{
        \includegraphics[width=0.185\linewidth]{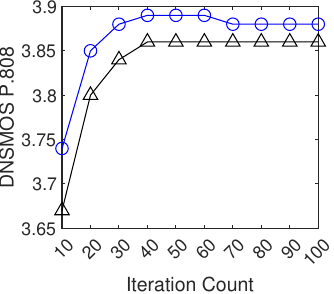}
        }
    \subfloat[WER with 'tiny' model]{
        \includegraphics[width=0.185\linewidth]{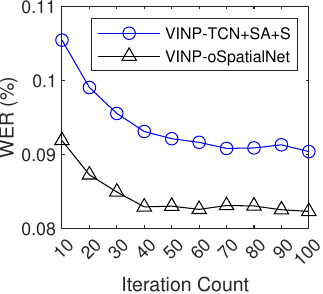}
        }
    
    \caption{\addnote[]{1}{Logarithmic likelihood and dereverberation metrics as a function of VEM iterations on REVERB SimData.}}
    \label{fig:CurveDereverb}
\end{figure*}
\subsection{Results and Analysis}
\subsubsection{Speech Dereverberation}

Empirically setting the number of VEM iterations to 100, the dereverberation results are presented in Table \ref{tab:results_SD}, where the bold font denotes the best results.
The results with negative effects are marked with a darkened background.

As the metrics of unprocessed speech and clean speech indicate, a larger ASR system implies stronger recognition ability and greater robustness to reverberation.
Compared with using a large ASR system alone, introducing an ASR-effective front-end speech dereverberation system can significantly improve the performance of ASR with fewer parameters and lower computational costs.
Among the comparison methods, only a few approaches are proven to be effective for ASR, such as GWPE, CMGAN, oSpatialNet*, and the proposed VINP-TCN+SA+S and VINP-oSpatialNet.
Among the DL-based methods, SkipConvNet, StoRM, and TCN+SA+S encounter failures in ASR, although some of them are effective on the 'tiny' model.
The reason for such ASR performance lies in the unpredictable artificial errors induced by DNN~\cite{iwamoto22_interspeech,iwamoto2024does,menne2019investigation}.
While these errors do not significantly affect speech perceptual quality, their impact on back-end speech recognition applications remains uncertain.
In contrast, larger ASR models are inherently more robust to noise and reverberation and place greater emphasis on extracting the intrinsic features of speech.
For these advanced models, however, the artificial distortions introduced by DNN-based dereverberation approaches become more pronounced, often resulting in a decline in recognition accuracy.
Different from the comparison methods, VINP employs a linear CTF filtering process to model the reverberation effect and leverages the DNN output as a prior distribution of anechoic source speech.
During the VBI procedure, VINP takes into account both this prior information and the observation.
By doing this, VINP bypasses the direct utilization of DNN output and ensures the outputs satisfy the signal model, consequently reducing the artifacts.
Meanwhile, VINP still utilizes the powerful non-linear modeling ability of DNN.
Consequently, VINP exhibits remarkable superiority over the other approaches in ASR performance.
\addnote[Ztest]{1}{When comparing VINP-TCN+SA+S with its backbone network TCN+SA+S, a gap in WER can be discerned, indicating that VINP can make an ASR-ineffective DNN effective. 
The WERs of oSpatialNet* and VINP-oSpatialNet on SimData with the 'medium' Whisper model are the same.
However, a gap still exists in other situations, indicating that VINP can make an ASR-effective DNN even more effective.
A Z-test is conducted between VINP and the backbones at a significance level of 0.05.
Results indicate that the differences in WER are significant, except for VINP-oSpatialNet and oSpatialNet* on the Whisper 'medium' model.
When comparing VINP w/o VEM with the original backbones, it becomes evident that the DNNs adapted to speech prior estimation in the proposed VINP yields a certain degree of performance improvement on ASR. 
However, such an improvement is still insufficient to make TCN+SA+S effective for ASR tasks.
Comparing VINP with VINP w/o VEM, it is clear that VEM further improves the ASR performance.}
Notably, VINP-oSpatialNet achieves SOTA in ASR.

\begin{figure}
    \centering
    \includegraphics[width=0.8\linewidth]{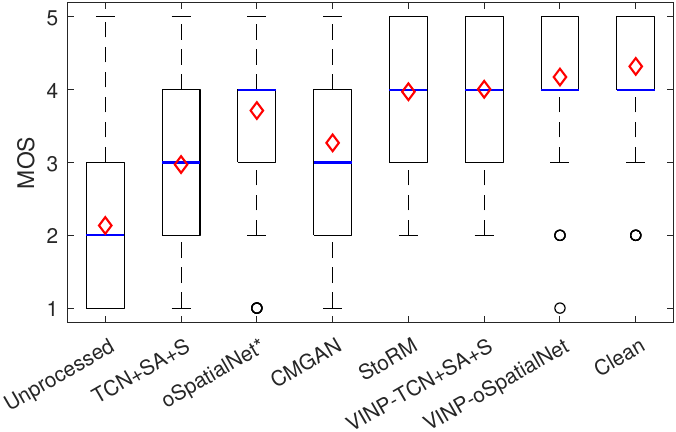}
    \caption{\addnote[]{1}{MOS of subject listening test.}}
    \label{fig:MOS}
\end{figure}

\begin{figure}[ht]
    \centering
    \subfloat[Reverberant speech]{
        \includegraphics[width=0.9\linewidth]{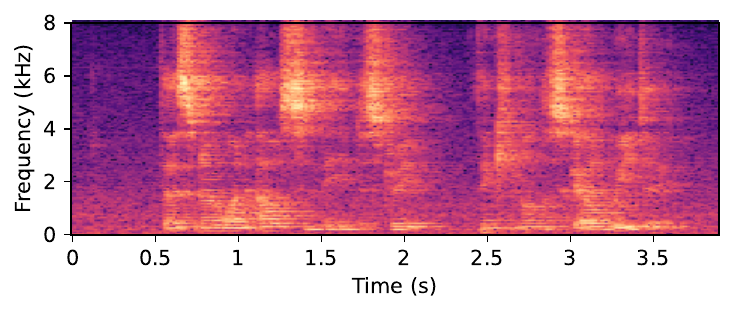}
        }\hfill
    \subfloat[Output of TCN+SA+S]{
        \includegraphics[width=0.9\linewidth]{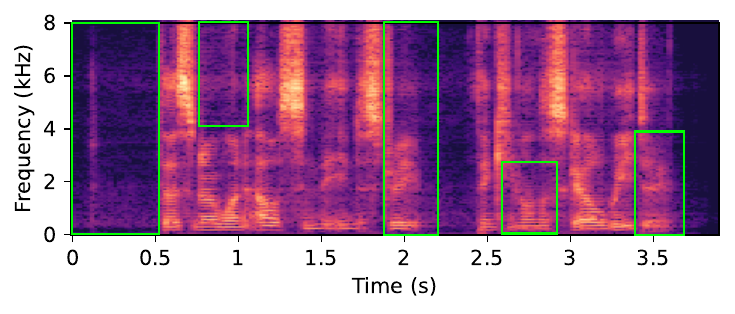}
        }\hfill
    \subfloat[Output of VINP-TCN+SA+S]{
        \includegraphics[width=0.9\linewidth]{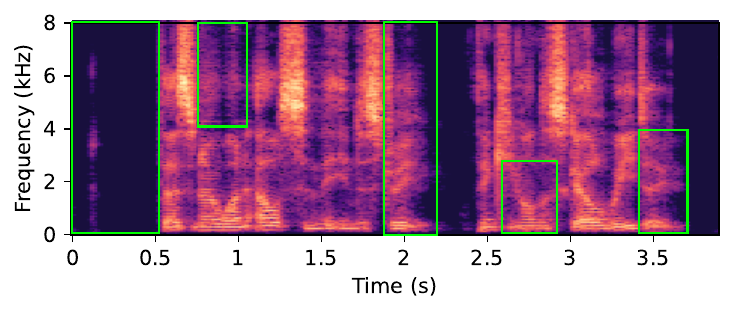}
        }\hfill
    \subfloat[Clean speech]{
        \includegraphics[width=0.9\linewidth]{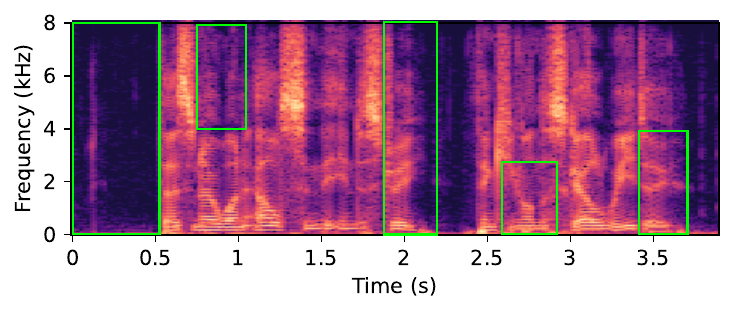}
        }\hfill
    \caption{Examples of magnitude spectra (RT60=0.7 s).}
    \label{fig:demo}
\end{figure}

Regarding speech quality, ESTOI of VINP is roughly the same as that of the backbone network, while PESQ shows a slight decrease compared to the backbone network.
\addnote[listeningtest]{1}{
As shown in Table \ref{tab:results_SD} and Fig. \ref{fig:MOS}, VINP-oSpatialNet achieves the highest MOS among all approaches.
Both VINP-oSpatialNet and VINP-TCN+SA+S exhibit higher MOS than their respective backbones, although their DNSMOS scores are close to those of TCN+SA+S.
MOS of CMGAN and StoRM also differ from their DNSMOS scores, indicating the limitations of DNSMOS as an evaluation metric for the present work.
Readers are encouraged to access the online audio examples for direct evaluation\footnote{\url{https://github.com/Audio-WestlakeU/VINP}}.
Comparing VINP with VINP w/o VEM, it can be found that VEM yields improvements across all perceptual metrics.
And the improvement of LSD shows that VEM brings improvements to the magnitude spectrum.}
Spectra of the reverberant speech, the output of TCN+SA+S, the output of VINP-TCN+SA+S, and the clean speech are shown in Fig.~\ref{fig:demo}.
Due to the insufficient modeling capability of the DNN, the output of TCN+SA+S contains overly smoothed artifacts.
On the contrary, VINP-TCN+SA+S not only refers to the priors provided by DNNs, but also directly utilizes the observation during the VBI procedure, resulting in a more accurate estimation of the anechoic speech spectrum.
Also, VINP-TCN+SA+S exhibits better performance in background noise control.

To illustrate the influence of VEM iterations on the performance of VINP, we show the average expected logarithmic likelihood of the complete data $\left<\ln p(\mathbf{S},\mathbf{X})\right>$ (where the constant is omitted) and some metrics on REVERB SimData in Fig. \ref{fig:CurveDereverb}.
As the iteration count increases, the logarithmic likelihood and all metrics shift in a positive direction. 
ESTOI, DNSMOS and WER tend to converge after 40 iterations.
In contrast, the logarithmic likelihood and PESQ show a continuous improvement.
This phenomenon indicates that better dereverberation performance can be achieved at the cost of computational complexity.
With 100 iterations, the logarithmic likelihood of VINP-TCN+SA+S is higher than that of VINP-oSpatialNet, yet the metrics of VINP-oSpatialNet are better. 
This indicates that when the DNN architecture varies, logarithmic likelihood cannot serve as an indicator to evaluate the enhancement performance of VINP.

\subsubsection{Blind RIR Identification}

Empirically setting the number of VEM iterations to 300, the RT60 and DRR estimation results are illustrated in Table \ref{tab:RIR_estimation}, where the bold font represents the best results.

Results show that VINP reaches the SOTA and advanced levels in RT60 and DRR estimation.
Theoretically, one limitation of our method is that a CTF filter with a finite length can only model a RIR with a finite length.
Nevertheless, it remains sufficient for RT60 and DRR estimation even under large-reverberation conditions, because the signal energy at the tail of the RIR is relatively negligible.
From the perspective of spatial acoustic parameter estimation, VINP can estimate RIRs well.
\begin{table}
\centering
\caption{\addnote[]{1}{Blind RT60 and DRR estimation results on SimACE}}
\label{tab:RIR_estimation}
\renewcommand\arraystretch{1.2}
\begin{tabular}{c|c|c|c|c}
    \Xhline{1pt}
    \multirow{2}{*}{Method}& \multicolumn{2}{c|}{RT60 (s)}& \multicolumn{2}{c}{DRR (dB)}\\
    \Xcline{2-5}{0.4pt}
    &MAE&RMSE&MAE&RMSE\\
    \Xhline{0.4pt}
    Jeub's~\cite{jeub2011blind}&-&-&7.14&8.69\\
    FiNS~\cite{steinmetz2021filtered}&0.113&0.167&\textbf{2.15}&\textbf{2.64}\\
    BUDDy~\cite{lemercier2024unsupervised}&0.124&0.173&3.93&4.57\\
    \Xhline{0.4pt}
    VINP-TCN+SA+S (prop.)&\textbf{0.089}&\textbf{0.124}&3.26&3.76\\
    VINP-oSpatialNet (prop.)&0.103&0.154&2.40&2.88\\
    \Xhline{1pt}
\end{tabular}
\end{table}
\begin{figure}
    \centering
    \subfloat[RT60 estimation]{
        \includegraphics[width=0.46\linewidth]{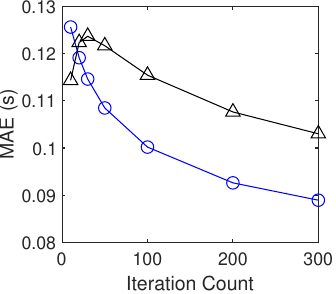}
        }
    \subfloat[DRR estimation]{
        \includegraphics[width=0.46\linewidth]{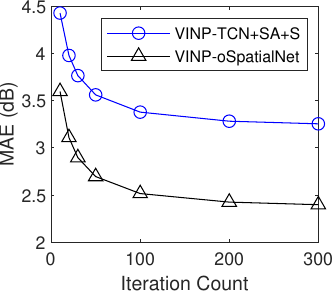}
        }
    \caption{\addnote[]{1}{MAE of RT60 and DRR estimation as a function of VEM iterations on SimACE.}}
    \label{fig:CurveRIR}
\end{figure}

We also show the MAE of RT60 and DRR estimation as the iteration count rises in Fig. \ref{fig:CurveRIR}.
As the iteration count increases, the errors of RT60 and DRR estimation decrease, indicating improved RIR identification.
It appears that the DRR tends to converge after 100 iterations, while the estimation accuracy of RT60 seems to still improve after 300 steps.
Therefore, the maximum number of iterations should be selected according to the objective.
VINP-oSpatialNet performs better in DRR estimation, while VINP-TCN+SA+S excels in RT60 estimation, which may be related to the DNN structure. 
Owing to the complexity of the entire system, we are currently unable to draw a conclusion regarding the preference for a specific DNN architecture in RIR identification.

\subsubsection{\addnote[]{1}{Noise Robustness}}
\begin{table}[H]
\centering
\caption{\addnote[]{1}{WER~(\%) with various noise levels and types}}
\label{tab:noise}
\renewcommand\arraystretch{1.2}
{
\begin{tabular}{c|c|c|c|c}
    \Xhline{1pt}
    \multirow{2}{*}{Method}&\multicolumn{2}{c|}{REVERB Noise}&\multicolumn{2}{c}{CHiME4 Noise}\\
    \Xcline{2-5}{0.4pt}
    &5 dB&20 dB&5 dB&20 dB\\
    \Xcline{1-5}{0.4pt}
    Unprocessed&30.0&13.3&26.8&12.9\\
    VINP-TCN+SA+S&21.7&9.0&-&-\\
    VINP-oSpatialNet&14.6&8.2&-&-\\
    Oracle prior + VEM&8.3&8.0&8.6&8.1\\
    \Xhline{1pt}
\end{tabular}}
\end{table}
\begin{table}[H]
\centering
\caption{\addnote[]{1}{PESQ with various noise levels and types}}
\label{tab:noise_pesq}
\renewcommand\arraystretch{1.2}
{
\begin{tabular}{c|c|c|c|c}
    \Xhline{1pt}
    \multirow{2}{*}{Method}&\multicolumn{2}{c|}{REVERB Noise}&\multicolumn{2}{c}{CHiME4 Noise}\\
    \Xcline{2-5}{0.4pt}
    &5 dB&20 dB&5 dB&20 dB\\
    \Xcline{1-5}{0.4pt}
    Unprocessed&1.15&1.48&1.11&1.43\\
    VINP-TCN+SA+S&1.79&2.52&-&-\\
    VINP-oSpatialNet&1.98&2.82&-&-\\
    Oracle prior + VEM&2.47&3.11&2.28&3.05\\
    \Xhline{1pt}
\end{tabular}}
\end{table}


\begin{table}[H]
\centering
\caption{\addnote[]{1}{MAE of RT60 estiamtion with various noise levels and types}}
\label{tab:noise_t60}
\renewcommand\arraystretch{1.2}
{
\begin{tabular}{c|c|c|c|c}
    \Xhline{1pt}
    \multirow{2}{*}{Method}&\multicolumn{2}{c|}{REVERB Noise}&\multicolumn{2}{c}{CHiME4 Noise}\\
    \Xcline{2-5}{0.4pt}
    &5 dB&20 dB&5 dB&20 dB\\
    \Xcline{1-5}{0.4pt}
    VINP-TCN+SA+S&0.406&0.089&-&-\\
    VINP-oSpatialNet&0.438&0.103&-&-\\
    Oracle prior + VEM&0.415&0.071&0.427&0.098\\
    \Xhline{1pt}
\end{tabular}}
\end{table}




\addnote[noiseExp]{1}{VINP is designed for applications under relatively simple noise conditions.
In this part, we adjust both the SNR and noise type in REVERB SimData and SimACE to demonstrate how the noise level and mismatched non-stationary noise affect the performance of VINP.
Specifically, we replace the stationary REVERB noise with that from CHiME4~\cite{thiemann2013diverse}, which was collected from four distinct environments: buses, cafes, pedestrian zones, and street intersections.
Most noise recordings in CHiME4 are non-stationary, and the transient noise is included.
Since there are already numerous DNN-based speech denoising methods, we do not discuss the influence of non-stationarity noise on the capability of DNNs, but use the oracle prior and focus on the effect of VEM.
The result using the oracle prior is the upper bound of the proposed framework.
WER using Whisper 'tiny' model and PESQ are shown in Table \ref{tab:noise} and Table \ref{tab:noise_pesq}, respectively.
The MAE of RT60 estimation is shown in Table \ref{tab:noise_t60}.
The results demonstrate that neither VINP-TCN+SA+S nor VINP-oSpatialNet achieves the performance ceiling due to errors in the DNN prediction.
As the SNR increases, the performance of both speech dereverberation and RIR identification tasks improves.
This is consistent with our expectations, as a higher SNR indicates a more reliable observation in the VEM procedure.
Compared to the results with matched stationary noise, the performance of VINP degrades under mismatched non-stationary noise conditions, particularly for the dereverberation task.}

\section{Conclusion}
\label{sec:conclusion}
In this paper, we propose a variational Bayesian inference framework with neural speech prior for joint ASR-effective speech dereverberation and blind RIR identification. 
Combining the neural prior distribution of anechoic speech with the reverberant recording, VINP employs VBI to solve the CTF-based probabilistic signal model and further estimate the anechoic speech and RIR.
The usage of VBI avoids the direct utilization of DNN output but still utilizes its powerful nonlinear modeling capability, and proves to be effective for ASR systems without any joint training.
Experimental results demonstrate that the proposed method achieves superior or competitive performance against the SOTA approaches in both tasks.

\ifCLASSOPTIONcaptionsoff
  \newpage
\fi


\bibliographystyle{IEEEtran}
\bibliography{bibtex/bib/IEEEexample}

\end{document}